\documentclass[12pt]{article}
\usepackage{amsmath}
\usepackage{amsfonts}
\usepackage{epsfig}
\newcommand{\Ref}[1]{(\ref{#1})}
\renewcommand{\phi}{\varphi}
\newcommand{\<}{\langle}
\renewcommand{\>}{\rangle}
\newcommand{\bpsi}{\bar\psi}

\title{\bf On the one-dimensional reggeon model:
eigenvalues of the Hamiltonian and the propagator.}
\author{M.A. Braun, E.M. Kuzminskii, A.V. Kozhedub, A.M. Puchkov,
 M.I. Vyazovsky\\
Dept. of High Energy physics,
Saint-Petersburg State University,\\
198504 S.Petersburg, Russia}

\begin{document}

\maketitle
{\bf Abstract}

The effective reggeon field theory in zero transverse dimension
(''the toy model'') is studied. The transcendental equation for eigenvalues
of the Hamiltonian of this theory is derived and solved numerically.
The found eigenvalues are used for the calculation of the pomeron propagator.

\section{The one-dimensional quantum reggeon model}

The effective theory of reggeon interaction (''the Gribov model'') was
introduced in \cite{gribov} as a set of diagrammatic rules with
an infinite number of vertices. In \cite{migdal,migdal1} by means of
the renormalization group method it was shown that
at high energies only the three-reggeon interactions
are important. In \cite{migdal} the diagrammatic rules were reconsidered
as Feynman rules for the Euclidean field theory in three dimensions with
the Lagrangian density
\begin{equation}
\mathcal{L}=\frac{1}{2} \left( \Phi^{+} \partial_y \Phi -
\Phi \partial_y \Phi^{+} \right) - \mu \Phi^{+} \Phi
-\alpha' \vec\nabla \Phi^{+} \cdot \vec\nabla \Phi
+ i \lambda \Phi^{+} \left( \Phi + \Phi^{+} \right) \Phi .
\label{e11}
\end{equation}
Here $\Phi(y,\vec b)$ -- the complex reggeon (pomeron) field,
$\Phi^{+}(y,\vec b)$ -- the conjugated field, $y$ -- rapidity parameter
(logarithm of total energy), $\vec b$ -- the two-dimensional impact
parameter. Parameters $\mu=\alpha(0)-1$ and $\alpha'=\alpha'(0)$ are
defined by the pomeron Regge trajectory $j=\alpha(t)$. The effective
coupling constant $\lambda$, because of physical reasons, has to be real
and positive. One has also to add field sources for $\Phi,\Phi^{+}$ which
describe the interaction of reggeons with scattering particles -- hadrons.
It is worth to note that theory \Ref{e11} arises as a continuous limit
in the two-dimensional directed bond percolation model \cite{cardy}.

At the same time, in a number of works
\cite{amati1,amati2,jengo,amati3,rossi}
a simplified one-dimensional model was considered
\begin{equation}
\mathcal{L}=\frac{1}{2} \left( \Phi^{+} \partial_y \Phi
- \Phi \partial_y \Phi^{+} \right) - \mu \Phi^{+} \Phi
+ i \lambda \Phi^{+} ( \Phi + \Phi^{+} ) \Phi,
\label{e12}
\end{equation}
where $\Phi,\Phi^{+}$ depend only on $y$.
The one-dimensional theory may be thought of as the lowest order
approximation of the three-dimensional theory with $\alpha' \to 0$,
since in this case the dynamic connection between fields with different
impact parameters $\vec b$ is lost. To a certain extent it can
correspond to the physical reality, since the slope of the pomeron
trajectory $\alpha'\!\sim\! 0.2 GeV^{-2}$ may be considered small
in comparison to the momenta scale in hadron scattering processes.
However, the one-dimensional model also serves as a ''toy model''
which allows to study the correlation between the results given
by the perturbative approach and non-perturbative methods.
Conclusions drawn from the analysis of this model can be significant
for studying more complicated and realistic theories, including
quantum chromodynamics. It is the one-dimensional model that is
considered in the presented work.

Theory \Ref{e12} describes a first-order formalism. Besides, this is an
Euclidean field theory in the sense that rapidity $y$ may be understood
as the imaginary time. The canonical quantization in imaginary time leads
to the following commutation relation
\begin{equation}
[\Phi , \Phi^{+}]=1
\label{e14}
\end{equation}
which allows one to interpret $\Phi^{+}$ as the creation operator and
$\Phi$ as the annihilation operator. The Hamiltonian has the form
\cite{rossi}
\begin{equation}
H=-\mu \Phi^{+}\Phi + i \lambda
\left( \Phi^{+}\Phi^{+}\Phi + i \lambda \Phi^{+}\Phi\Phi \right) .
\label{e13}
\end{equation}
Following \cite{rossi}, we use a normal order of operators. For the
Hamiltonian \Ref{e13} the existence of the Fock space is postulated.
Quantum field theory \Ref{e12} without spatial dimensions is equivalent
to the one-dimensional quantum mechanics, at least in the perturbation
(in the coupling constant) theory framework. It is important that
the perturbation theory works well for $\mu<0$, when eigenvalues
of the unperturbed Hamiltonian are bounded from below.

In work \cite{schwimmer} this theory was considered in the ''loopless''
approximation, when one of the terms inside brackets in \Ref{e13} is
ignored. In this case the theory becomes integrable. However, in
\cite{braun1,braun2} the evolution in rapidity of the hadron scattering
amplitude in the framework of the one-dimensional reggeon model was studied
numerically and it was shown that the difference between the full theory
and the ''loopless'' approximation becomes substantial at high $\lambda$.
Thus it is meaningful to develop methods that are not connected with
the perturbation theory in $\lambda$.

In \cite{bondarenko} the eigenvalues of the Hamiltonian \Ref{e13}
were found for values of $\mu/\lambda=1,3,5$ and the propagator
at $\mu/\lambda=5$ was calculated, although neither the calculational
procedure nor the precision were reported. In this paper we concentrate
on the theoretical problems related to the Hamiltonian in the complex
plane and on the method of calculation of its eigenvalues and
eigenfunctions. We enlarge the domain of the values of $\mu/\lambda$
to include negative ones for comparison with the perturbative approach.

\section{Eigenvalues problem}

In this section we mostly reproduce the results of \cite{rossi}
which serve as a starting point of our study.
It is convenient to use the Bargmann representation for the Fock space
in the form of the Hilbert space of analytical functions, in order to use
methods from the differential equations theory. This representation
is introduced by the relation
\begin{equation}
\psi(z)=\langle 0 | e^{\Phi z} | \psi \rangle
\label{e21}
\end{equation}
for a wave function. By construction $\psi(z)$ is an analytical function
in $z$. In the Bargmann representation multiplication and differentiation
correspond to the creation and annihilation operators
\begin{equation}
\Phi^{+} \to z ,\quad \Phi \to \frac{d}{dz}\ ,
\label{e22}
\end{equation}
the basis Fock state $|n\> = (\Phi^{+})^n |0\>$
then corresponds to $\psi(z)=z^n$.
The standard scalar product of the Fock space in the frame of analytical
functions $\psi(z)$ is reproduced by the formula
\begin{equation}
\< \psi_1 | \psi_2 \> =
\int \frac{dz dz^*}{2 \pi i} e^{-zz^*} \psi_1^*(z) \psi_2(z).
\label{e23}
\end{equation}

Different basis states are orthogonal, what is easily seen for
$\psi_2(z)=z^n$ and $\psi_1^*(z)=(z^*)^m$, if one carries in
\Ref{e14} the integration over $\rm{arg}(z)$. The basis state norm
is defined by the integral over absolute value of $z$ (here $t=|z|^2$):
\begin{equation}
<n|n>= \int_0^{\infty}\!d|z|\cdot 2|z|\  e^{-|z|^2}|z|^{2n}
=\int_0^{\infty}\!dt\  e^{-t}t^{n} = \Gamma(n+1) = n! \ .
\label{e24}
\end{equation}
Hamiltonian \Ref{e13} in the Fock-Bargmann representation takes the form
\begin{equation}
H= -\mu z \frac{d}{dz}+
i \lambda z \frac{d^2}{dz^2} + i \lambda z^2 \frac{d}{dz} .
\label{e25}
\end{equation}

We want to find eigenvalues and eigenstates of the Hamiltonian. The equation
\begin{equation}
H\psi=E\psi
\label{e27}
\end{equation}
is a linear differential equation of the second order (of the Heun class,
see the next section) the general solution of which is a linear combination
of two linearly independent functions. Their asymptotical behaviour at $z=0$
are found to be
\begin{equation}
\psi_1(z) \underset{z \to 0}{\sim} z \quad \mbox{or} \quad 
\psi_2(z) \underset{z \to 0}{\sim} c_1 + c_2 z \ln z.
\label{e28}
\end{equation}
Because of analyticity of $\psi(z)$, it has to be $c_2=0$, which
is possible only when $E=0$. In this case the normalizable in the
Fock-Bargmann space solution $\psi_0(z)=1$ is the vacuum
wave function, which has no physical meaning for reggeon theory (from
this point of view the state $|0\>$ describes the absence of interaction
of physical particles with reggeon). Therefore, the second asymptotics
is to be excluded.

The asymptotics in the vicinity of infinity is also known
\begin{equation}
\psi_3(z) \underset{z \to \infty}{\sim} const, \quad
\psi_4(z) \underset{z \to \infty}{\sim} 
\frac{1}{z} e^{- \frac{1}{2} z^2 - \frac{i \mu}{\lambda}z} .
\label{e29}
\end{equation}
It is worth to note that the asymptotical behaviour \Ref{e29}
depends on the direction in the complex plane.

Let us begin with the case $\mu<0$, which corresponds to the domain of the
applicability of the perturbation theory. As in \cite{rossi} we impose
the condition of a finite Bargmann norm.
Since asymptotics of the integrand in \Ref{e23} for $\<\psi_4|\psi_4\>$ is
\begin{equation}
e^{-|z^2|} \left| \frac{1}{z} e^{- \frac{z^2}{2}
- \frac{i \mu}{\lambda} z} \right|^2 \simeq
\frac{1}{|z|^2}
e^{\frac{2 \mu}{\lambda} \mathrm{Im}\, z} e^{-2 (\mathrm{Re}\, z)^2},
\label{e210}
\end{equation}
the integral \Ref{e23} can not converge on $\mathrm{Im}\, z$ if a wave
function has asymptotics $\psi_4$ on the negative imaginary axis $z$
(it is always considered that $\lambda>0$). A solution with asymptotics
$\psi_3$ in this direction will have a finite norm. Thus the quantization
condition is $\psi\sim\psi_3$, when $z\to -i\infty$.

For further convenience the following substitution will be used 
\begin{equation}
z=-i\sqrt{2}x \ .
\label{e220}
\end{equation}
The eigenfunctions equation $H\psi=E\psi$ may be rewritten
(for $\lambda\neq 0$) in the form
\begin{equation}
x \psi '' + \left( \frac{\mu \sqrt{2}}{\lambda}x - 2x^2 \right) \psi'
+ \frac{E \sqrt{2}}{\lambda} \psi=0,
\label{e221}
\end{equation}
where a prime, here and everywhere further, denotes a derivative with
respect to $x$. Asymptotical boundary conditions for solution $\psi(x)$,
analytical in the entire complex plane $x$,
have to be set on the ray $x \in [0, +\infty)$:
\begin{equation}
\psi \underset{x \to 0}{\sim} x, \quad
\psi \underset{x \to +\infty}{\sim} const.
\label{e222}
\end{equation}

Note that in the case $\mu>0$, the finite Bargmann norm condition
$\psi\sim\psi_3$ has to be put on the positive imaginary axis $z$,
i.e. on the ray $x \in (-\infty, 0]$. After substitution $x\to -x$
the equation takes the form
\begin{equation}
x \psi '' + \left( \frac{- \mu \sqrt{2}}{\lambda}x - 2x^2 \right) \psi'
+ \frac{- E \sqrt{2}}{\lambda} \psi=0
\label{e227}
\end{equation}
with the same boundary conditions \Ref{e222}. It is equivalent to the
solution of the equation \Ref{e221} with $\mu\to -\mu<0$, $E\to -E$.
Calculations made in Section 5 show that for $\mu<0$ all values $E$ are
positive. Therefore solutions of the equation \Ref{e227} correspond to
unphysical negative values of $E$. This leads, e.g., to the infinite
growth of the propagator (see Section 6) when $y \to +\infty$. Thus
such formulation of the problem seems to be incorrect in the case $\mu>0$.
In Section~5 for $\mu>0$ the same conditions \Ref{e222} for the equation
\Ref{e221} on eigenfunctions were used. As a result, all found eigenvalues
$E$ are positive.

When $\mu=0$, the finiteness of the norm does not exclude both
asymptotics $\psi\sim\psi_{3,4}$ on the ray $x \in [0, +\infty)$
and on the ray $x \in (-\infty, 0]$. In this case the conditions
\Ref{e222} can be used as well, excluding asymptotics $\psi_4$,
and all the eigenvalues are found to be positive. Furthermore, in Section~6
the completeness property of the found eigenfunctions was partially
checked, and as a result, the necessary solutions were not lost.

In \cite{rossi} it was proven that the spectral representation
of the $S$-matrix of the theory \Ref{e13} exists and it is analytical
in $\mu$ on the entire real axis $\mu$. The choice of quantization
conditions \Ref{e222} for all values of $\mu$ is to be understood
in the sense of such analytical continuation. The positiveness of
the found eigenvalues indicates the correctness of this approach.

Hamiltonian \Ref{e25} is non-Hermitian with respect to the standard scalar
product \Ref{e23}. Hence a question arises, whether the eigenfunctions
defined by the finite norm condition form the complete basis. The answer
is given by the transformation procedure of \Ref{e25} to the Hermitian form,
proposed in \cite{jengo}. On the negative imaginary
axis $z=-iq$, $q=\sqrt{2}x>0$
\begin{equation}
H \equiv \lambda q \left( -\frac{d^2}{dq^2}
+ \left( q - \frac{\mu}{\lambda} \right) \frac{d}{dq} \right) .
\label{e223}
\end{equation}
By means of a similarity transformation $\tilde{H}=e^{-F(q)} H e^{F(q)}$
with $F(q)=e^{\frac{1}{4} \left( q - \frac{\mu}{\lambda} \right)^2}$
the term with a first-order derivative may be annihilated. After
substitution $q=\xi^2$, $\xi>0$ and one more transformation
\begin{equation}
H_{\xi} = \xi^{-\frac{1}{2}} \tilde{H} \xi^{\frac{1}{2}}
=\frac{\lambda}{4} \left( -\frac{d^2}{d\xi^2} + \frac{3}{4\xi^2}
+ \xi^2 \left(\left( \xi^2 - \frac{\mu}{\lambda} \right)^2 -2 \right)\right)
\label{e224}
\end{equation}
takes the form of a Hermitian Hamiltonian with a singular potential.
Similarity transformation from $H$ to $H_{\xi}$ is non-unitary,
but bijective. Equation $H\psi=E\psi$ is then equivalent to
$H_{\xi}\phi=E\phi$, where
\begin{equation}
\phi(\xi)=\xi^{-\frac{1}{2}}
e^{-\frac{1}{4} \left( \xi^2 - \frac{\mu}{\lambda} \right)^2}
\psi(z=-i\xi^2)
\label{e225}
\end{equation}
is an analytical function in $\xi$ in the vicinity of half-axis
$(0, + \infty)$ and finite when $\xi \to 0$. The choice
of functions $\psi(z)\underset{z \to 0}{\sim} z$ means that
$\phi(\xi)\underset{\xi \to 0}{\sim} \xi^{3/2}$ -- this behaviour
corresponds to the angular momentum barrier for the potential
in \Ref{e224}.

The asymptotics of eigenfunctions at $\xi\to +\infty$,
corresponding to \Ref{e29}, are defined by the relations
\begin{equation}
\phi_3(\xi) \sim \xi^{-\frac{1}{2}}
e^{-\frac{1}{4} \left( \xi^2 - \frac{\mu}{\lambda} \right)^2},
\quad
\phi_4(\xi) \sim \xi^{-\frac{5}{2}}
e^{\frac{1}{4} \xi^4 - \frac{\mu}{2\lambda} \xi^2} .
\label{e226}
\end{equation}
The finiteness of the norm of $\phi(\xi)$ in the space $L_2([0, + \infty))$
excludes the second asymptotics, what is equivalent to the choice of
asymptotics $\psi_3$ for $\psi(z)$, i.e. the Bargmann norm finiteness. Thus
previously formulated problem of finding the spectrum of Hamiltonian $H$
is equivalent to the problem of finding the spectrum of Hermitian
operator $H_{\xi}$ in the space $L_2([0, + \infty))$. Eigenfunctions of
this problem are a complete basis; all the eigenvalues, common for $H$
and $H_{\xi}$, are real. Note that the Hermiticity of $H_{\xi}$ does not
depend on the sign of $\mu$.

\section{Biconfluent Heun equation}

The canonical form of the biconfluent Heun equation is
($\alpha,\beta,\gamma,\delta$ are constant parameters)
\begin{equation}
x\psi''(x) + \left( 1+\alpha-\beta x - 2 x^2 \right) \psi'(x)
+ \left( (\gamma - \alpha - 2) x 
  -\frac{1}{2}(\delta + (1 + \alpha )\beta ) \right) \psi(x)=0 .
\label{e31}
\end{equation}
By means of the transformation
\begin{equation}
u(x)=x^{\frac{1+\alpha}{2}}
\exp{\left( - \frac{\beta}{2}x - \frac{1}{2}x^2 \right)} \psi(x)
\label{e32}
\end{equation}
equation \Ref{e31} is brought into the so-called normal form
\begin{equation}
u''(x) + \left( \frac{1}{4} \left( 1-\alpha^2 \right) \frac{1}{x^2}
- \frac{1}{2} \delta \frac{1}{x} + \gamma - \left( \frac{\beta}{2} \right)^2
- \beta x - x^2 \right) u(x)=0 .
\label{e33}
\end{equation}
Equation \Ref{e221}, rewritten as
\begin{equation}
x\psi''(x) + ( -\beta x -2 x^2 ) \psi'(x)
-\frac{\delta}{2} \psi(x)=0 ,
\label{e34}
\end{equation}
is the canonical form of biconfluent Heun equation with the parameters
\begin{equation}
\alpha=-1, \quad \beta=-\frac{\sqrt{2}\mu}{\lambda}, \quad \gamma=1 ,
\quad \delta=-\frac{2\sqrt{2}E}{\lambda} .
\label{e35}
\end{equation}

Behaviour of solutions of the equation \Ref{e31} is well known
\cite{slavyanov}. Let us remind main facts about it (useful formulae may
also be found in \cite{numalg}). The equation has two singular points,
a regular one at $x=0$ and an irregular one at $x=\infty$ with
the rank of singularity $R=3$. A solution of the equation can not have
any finite singularities apart from $x=0$.

Generally, the leading asymptotics near zero is a power
$x^{\frac{\alpha\pm 1}{2}}$. Our case $\alpha=-1$ is special in the sense
that only one of the solutions can be represented in the form of power
series (Frobenius series)
\begin{equation}
\psi(x)= \sum_{n=0}^{\infty} c_n x^{n+1} ,
\label{e38}
\end{equation}
converging in the entire complex plane. Its coefficients are defined
(up to a common factor) by the three-term recurrence
relation
\begin{equation}
n(n+1) c_{n}
=\left( n \beta + \frac{\delta}{2} \right) c_{n-1} + 2(n-1) c_{n-2} ,
\quad c_0 =1.
\label{e39}
\end{equation}
We will always consider $c_0 =1$ for solutions.
Thus $\psi(x)\sim x$ when $x \to 0$.
The second linearly independent solution with $\alpha=-1$ has the form
\begin{equation}
\tilde{\psi} (x)=\psi(x) \ln x + f(x),
\label{e310}
\end{equation}
where $f(x)$ is analytical function in $x$, and has to be excluded, because
it has a logarithmic branch point $x=0$. Only the solutions of the type
\Ref{e38}, analytical in the entire complex plane $x$, have to be considered.

In the vicinity of $x=\infty$ one can choose as two linearly independent
solutions of the equation \Ref{e34} the functions $\psi_3$ and $\psi_4$,
defined by the asymptotic power series (Thom\`e series):
$$
\psi_3(x)
\underset{x \to \infty}{\sim}
\sum_{n=0}^{\infty} a_n^3 x^{-n},
$$
\begin{equation}
\psi_4(x)
\underset{x \to \infty}{\sim}
\frac{1}{x} \exp\left( x^2+\beta x \right)
\sum_{n=0}^{\infty} a_n^4 x^{-n} .
\label{e311}
\end{equation}
As can be seen, asymptotics \Ref{e29} are the first terms of regular
expansions \Ref{e311}. The coefficients of these expansions are fully
defined by the recurrence relations
$$
2n a^3_n= \left( -(n-1)\beta +\frac{\delta}{2} \right) a^3_{n-1}
- (n-1)(n-2) a^3_{n-2} ,
$$
\begin{equation}
2n a^4_n= -\left( n \beta + \frac{\delta}{2} \right) a^4_{n-1}
+ n(n-1) a^4_{n-2} ,
\label{e312}
\end{equation}
where again we fix $a^{3,4}_0 =1$. Asymptotics of a general solution of
the equation \Ref{e34} is a linear combination of $\psi_3$ and $\psi_4$.

It is known \cite{slavyanov} that the asymptotics, defined by Thom\`e
series, is not reached uniformly in $\mbox{arg}(x)$. In our case, when the
irregular singular point $x=\infty$ has the rank $R=3$, there exist $R+1=4$
Stokes rays $\mbox{arg}(x)=\pi k/4$ ($k=0,1,2,3$), connecting
singular points $0$ and $\infty$ and dividing complex plane $x$
in 4 sectors. In the general case, in any of these sectors and on any
of Stokes rays the asymptotics of the solution may be different.
In the Appendix it is shown that asymptotics $\psi\sim const$ is reached
in the area $-\frac{\pi}{2} < \mbox{arg}(x) < \frac{\pi}{2}$,
i.e. on the ray $[0, + \infty)$ and two adjacent sectors.

\section{Orthogonal scalar product}

One of the features of a non-Hermitian Hamiltonian is
that its eigenstates $\psi_N$, corresponding to different $E_N$, are not
orthogonal with respect to the usual scalar product, even if $\psi_N$ form
the complete basis. If we introduce, following \cite{rossi}, the operator
$P$, changing the sign of $z$,
\begin{equation}
P z P^{-1} = -z , \quad P \frac{d}{dz} P^{-1} = -\frac{d}{dz} ,
\label{e41}
\end{equation}
then it is easily seen that
\begin{equation}
H^{+} =H[-\lambda] =P H P^{-1} .
\label{e42}
\end{equation}
The operator $P$ is unitary and Hermitian in the original Fock space,
since $\<P\psi_1|P\psi_2\>\equiv \<\psi_1|\psi_2\>$ and
$\<\psi_1|P\psi_2\>\equiv \<P\psi_1|\psi_2\>$ for the Bargmann scalar
product. One can define biorthogonal eigenvectors
\begin{equation}
\bpsi_N=-P\psi_N=-\psi_N(-z)
\label{e43}
\end{equation}
(sign ''-'' is convenient for formulae in Section 6).
If $H\psi_N=E_N \psi_N$ with real $E_N$, then
\begin{equation}
H^{+} \bpsi_M = - H^{+} P\psi_M = - P H \psi_M = E_N \bpsi_M .
\label{e44}
\end{equation}
Using this and the adjoint equation $\<\bpsi_M|H = E_M \<\bpsi_M|$,
one obtains
\begin{equation}
(E_N-E_M) \<\bpsi_M|\psi_N\> = \<\bpsi_M|(H-H)|\psi_N\> = 0 ,
\label{e45}
\end{equation}
so that $\<\bpsi_M|\psi_N\>=0$ when $E_N \neq E_M$.

The scalar product
$$
(\psi_1|\psi_2) = -\<\psi_1|P|\psi_2\> \equiv -<\psi_1(-z)|\psi_2(z)>
$$
\begin{equation}
= -\int \frac{dz dz^*}{2 \pi i} e^{-zz^*} \psi_1^*(-z) \psi_2(z) ,
\label{e46}
\end{equation}
has an orthogonality property, but is, in the general case, not positively
defined. Evidently, defining the Hermitian operator of sign $\nu$
on the basis states
$$
\nu\psi_N=\nu_N \psi_N , \quad \mbox{where}
$$
\begin{equation}
\left\{
\begin{array}{cc}
\nu_N=+1, & \mbox{if}\ \<\psi_N|P|\psi_N\> >0, \\
\nu_N=-1, & \mbox{if}\ \<\psi_N|P|\psi_N\> <0,
\end{array}
\right.
\label{e47}
\end{equation}
it is possible to introduce $\bpsi_N=\nu P\psi_N$ and the positively
defined scalar product (as in \cite{rossi})
$(\psi_1|\psi_2) = \<\psi_1|\nu P|\psi_2\>$,
with respect to which eigenstates are orthogonal. Note that
$\<\psi_N|P|\psi_N\>=0$ is impossible because it contradicts the
completeness of the set of $\psi_N$. The resolution of the identity
then takes the form \cite{nonherm1,nonherm2}
\begin{equation}
I=\sum_N \frac{|\psi_N><\bpsi_N|}{<\bpsi_N|\psi_N>} \ .
\label{e48}
\end{equation}
However, this relation does not change if $\psi_N$ and $\bpsi_N$ are
multiplied on an arbitrary constant, so it is not necessary to choose
the scalar product positively defined. For the sake of simplicity, the
definition \Ref{e43} will always be used.

The orthogonal scalar product may be defined in different ways
\cite{nonherm1,nonherm2}. For example, the standard scalar product of
$L_2([0, +\infty))$ with the help of the transformation \Ref{e225}
induces the following scalar product
\begin{equation}
(\psi_1|\psi_2)= \<\phi_1|\phi_2\>
=\int_{0}^{+\infty}\!d\xi\ \xi^{-1}
e^{-\frac{1}{2} \left( \xi^2 - \frac{\mu}{\lambda} \right)^2}
\psi_1^*(z=-i\xi^2) \psi_2(z=-i\xi^2).
\label{e49}
\end{equation}
Since Hamiltonian $H_{\xi}$ is Hermitian, its eigenstates $\phi_N$ with
different $E_N$ are orthogonal, so connected to them by \Ref{e225}
states $\psi_N$ are orthogonal with respect to \Ref{e49}. The inconvenience
of this definition is that with respect to it basis states $|n\>$ are not
orthogonal, what makes the norm calculation for wave functions, written
in the form \Ref{e38}, very cumbersome.

\section{Numerical calculation of eigenvalues: the method and results}

\begin{figure}[t]
\hskip 3cm
\epsfig{file=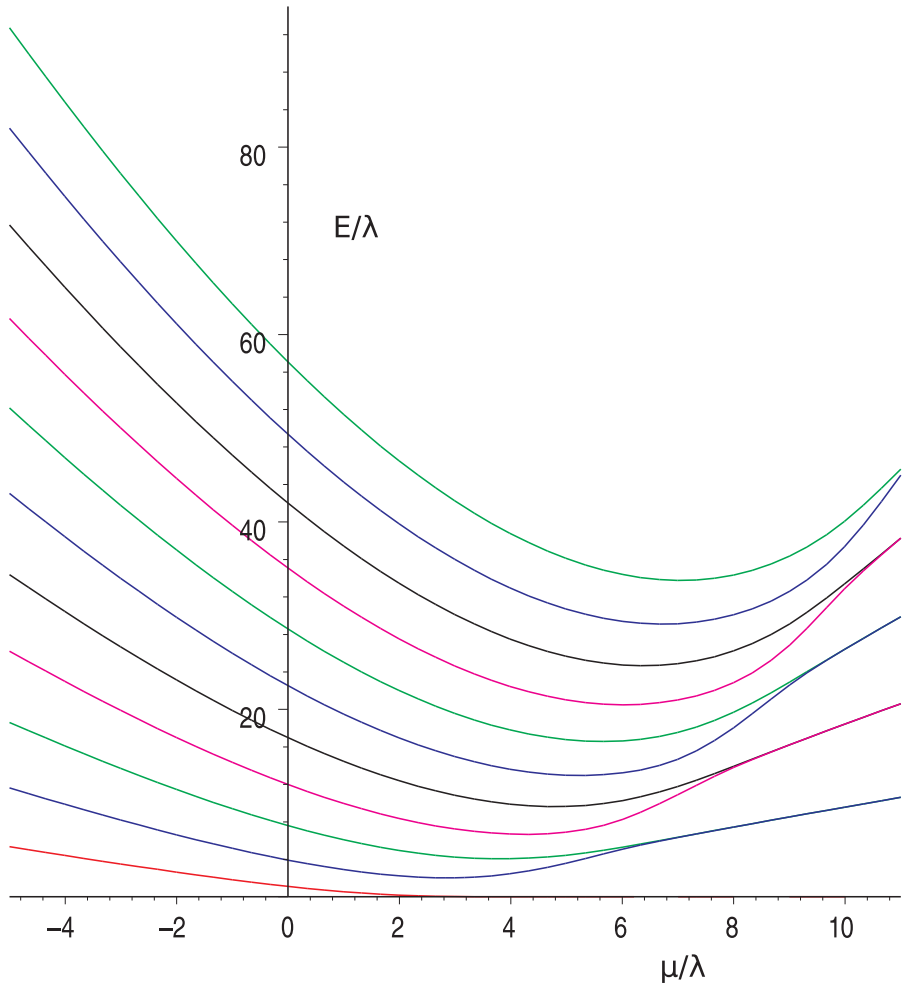,scale=0.9}
\caption{Plot of the universal ratio $E/\lambda$
for the first 11 eigenvalues as functions
of $\mu/\lambda$ from $-5$ to $+11$.}
\label{eigen11}
\end{figure}

\begin{figure}[t]
\hskip 3cm
\epsfig{file=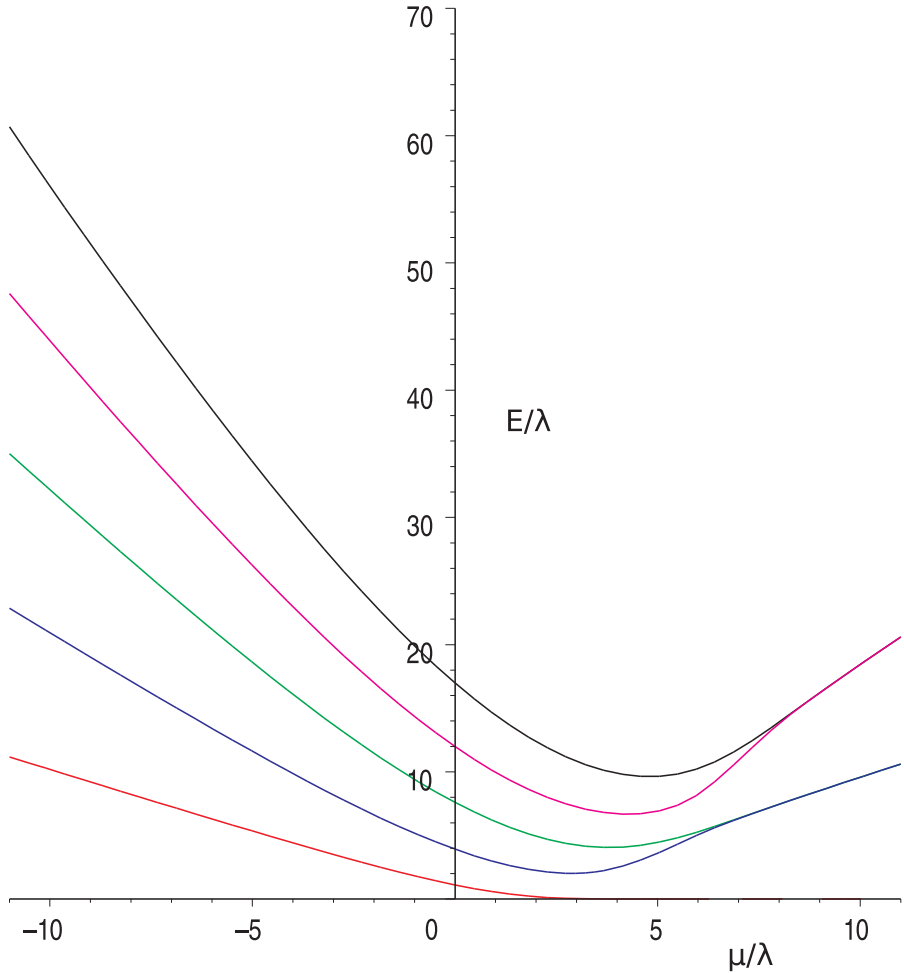,scale=0.9}
\caption{Plot of the universal ratio $E/\lambda$
for the first 5 eigenvalues as functions
of $\mu/\lambda$ from $-11$ to $+11$.}
\label{eigen5}
\end{figure}

Any solution \Ref{e38} of the equation \Ref{e34} behaves at infinity as
\begin{equation}
\psi(x) \underset{x \to \infty}{\sim} T_{3} \psi_3(x) + T_{4} \psi_4(x).
\label{e51}
\end{equation}
Here constants $T_{3}$ and $T_{4}$ called connection factors are
functions of two parameters $\beta$ and $\delta$ of \Ref{e34}. These
constants $T_{3,4}$ do not depend on $x$ but can differ in different
sectors of the complex plane which are divided by Stokes rays.
The expression for $T_{4}$ obtained in the Appendix is valid
on the half-plane $-\frac{\pi}{2} < \mbox{arg}(x) < \frac{\pi}{2}$.

It follows from the consideration of Sec. 2 that $T_{4}=0$ is the
quantization condition. For $\mu$ and $\lambda$ fixed and so for fixed
$\beta=-{\sqrt{2}\mu}/{\lambda}$ the equation $T_{4}=0$ has to be solved
for the variable $\delta=-{2\sqrt{2}E}/{\lambda}$. From the equation
\Ref{e34} and definitions \Ref{e35} it is evident that the ratio
$E/\lambda$ can depend only on the ratio $\mu/\lambda$. Then all
eigenvalues $E_N$ are expressed in the form
\begin{equation}
E_N= - \lambda \frac{\delta_N(\mu/\lambda)}{2\sqrt{2}} ,
\label{e52}
\end{equation}
where $\delta_N$ are all solutions of the equation $T_{4}=0$.

In principle, all connection factors can be found using the general method
for differential equations of the Heun class described in \cite{slavyanov}.
However, this method seems to be very complicated. For this reason we use
a comparatively new and more simple method proposed by the authors of
\cite{numalg} for the biconfluent Heun equation and based on the asymptotic
series for Wronskians. It allows one to obtain the value of $T_{4}$
in the form of a transcendental expression
\begin{equation}
T_{4}= -\frac{1}{2}\left(
\frac{\Gamma (\mathcal{N}+\frac{3}{2})}
     {\left( \frac{1}{2} \right)^{\mathcal{N}+\frac{1}{2}}}
  g_{2\mathcal{N}}
+\frac{\Gamma (\mathcal{N}+2)}
      {\left( \frac{1}{2} \right)^{\mathcal{N}+1}}
  g_{2\mathcal{N}+1}
\right) ,
\label{e53}
\end{equation}
where $\mathcal{N}$ is an arbitrary positive integer and (see \Ref{e911})
\begin{equation}
g_s=\sum_{n=0}^{\infty} a^3_n
\left( -\hat{c}_{s+n-1} -\beta \hat{c}_{s+n}
-(s+2n+2)\hat{c}_{s+n+1} \right).
\label{e54}
\end{equation}
The details of the method are described in the Appendix, the definition
of $\hat{c}_{n}$ is also given there, $a^3_n$ is defined in \Ref{e312}.
The left side of the equation $T_{4}=0$ contains the infinite series
in \Ref{e54} with terms expressed via coefficients which are defined
by recurrence relations. It is not clear, if one can sum the series
analytically, so we have used \Ref{e53} for the numerical calculation
of $\delta$ for the chosen set of ratios $\mu/\lambda$.

The numerical solutions of the equation $T_{4}=0$ were found\footnote{All
numerical calculations and graphical plotting were carried out using
{\tt Maple} computer algebra system.} for all integer values of
$\mu/\lambda$ from $-11$ to $+11$. These values were chosen to compare
our solutions with the results of \cite{braun1}. To check the smooth
dependence of solutions on $\mu/\lambda$ in the vicinity
of the peculiar value $\mu=0$ we also found the solutions for
$\mu/\lambda=\pm 0.1, \pm 0.25, \pm 0.5$.

For the numerical calculation one can take a finite number $\mathcal{T}$
of terms in \Ref{e54}, thus $T_{4}$ becomes a polynomial in $\delta$.
The convergence of solutions was checked by comparing roots found for
$\mathcal{T}-50$ and $\mathcal{T}$ terms in the equation. It is worth
to note that to find roots with large absolute values one has to increase
simultaneously the number of terms $\mathcal{T}$ and the precision
of calculations. We chose the limit of absolute values of roots
correspondent to $|E/\lambda|<99$. For this it is sufficient to take
$\mathcal{T}=400$ and to use the precision of 55 digits in the calculation.
The independence of roots on the arbitrary number $\mathcal{N}$ was
not checked, we always set $\mathcal{N}=50$. However, we checked that for
all considered $\mu/\lambda$ the equation has no real positive
roots $\delta$ of the same order correspondent to negative energies.
Positive roots with much larger values and also complex roots may
appear as an artefact of finiteness of $\mathcal{T}$.

Plots of the universal ratios $E_N/\lambda$ for the first 11 eigenvalues
(interpolated by cubic splines) are shown in Fig.~\ref{eigen11} for
$\mu/\lambda$ from $-5$ to $+11$, plots in Fig.~\ref{eigen5} are
for the first 5 eigenvalues for $\mu/\lambda$ from $-11$ to $+11$.
These plots demonstrate the approximate double degeneracy of
eigenvalues at $\mu/\lambda\to +\infty$ noted in \cite{rossi}.
In the data files applied to this article one can find
non-trivial ($E\neq 0$) ratios $E/\lambda$ with 20-digit
precision.

The method of \cite{numalg} was proposed recently and its applicability
has not been supported enough. Therefore we checked our found eigenvalues
by the usual ''shooting'' method. This test shows a full agreement of the
results obtained by both methods on the level of the adopted accuracy.
We plan to devote a separate publication to the comparative analysis
of different calculational algorithms for this eigenvalue problem.

\begin{figure}[h]
\hskip 4.7cm
\epsfig{file=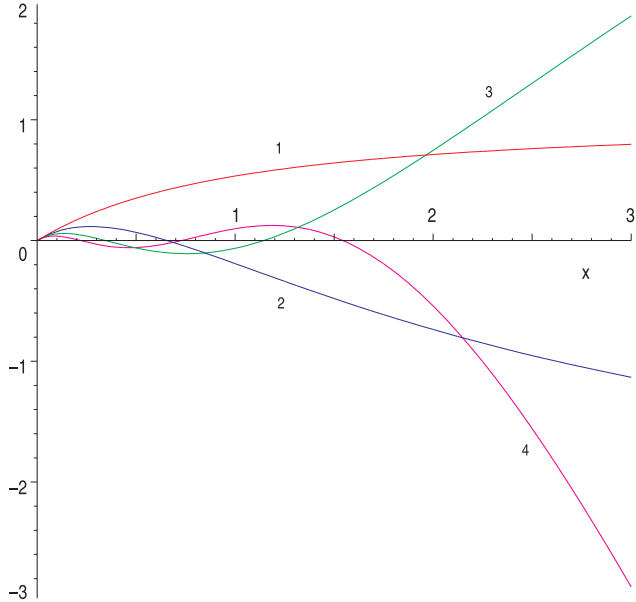,scale=0.6}
\caption{Plot of the first 4 eigenfunctions for the case
$\mu=0.1$, $\lambda=1$ at small $x$.}
\label{functions4}
\end{figure}

With the known value of $\delta$ correspondent to the eigenvalue $E_N$
one can find all coefficients of the series \Ref{e38} for the
eigenfunction from the recurrence relation \Ref{e39}. Since values
$\beta$ and $\delta$ determine the analytical eigenfunction uniquely,
any eigenvalue $E_N$ cannot be exactly degenerate. For example,
in Fig.~\ref{functions4} plots of the first 4 eigenfunctions are shown
for the ''almost physical'' case $\mu=0.1$, $\lambda=1$. Note that
for small values $x<3$ the asymptotic behaviour $\psi\sim const$
is not reached yet. These eigenfunctions have the usual properties
-- the $N$-th eigenfunction has $N$ zeroes on the real positive axis
including the point $x=0$. Obviously, the trivial vacuum eigenfunction
$\psi_0=1$ with $N=0$ has no zeroes.

\begin{figure}[h]
\hskip 4.7cm
\epsfig{file=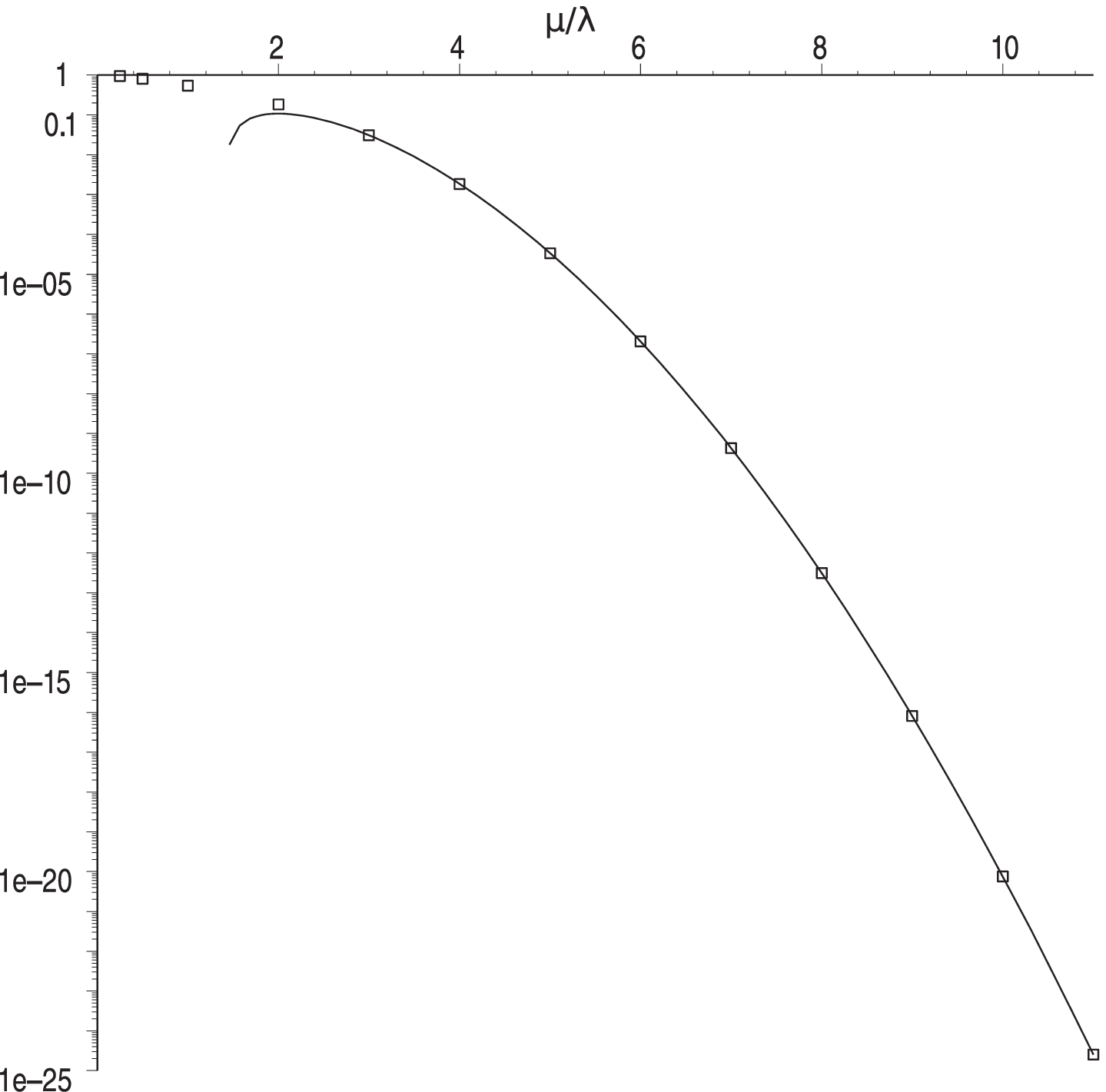,scale=0.6}
\caption{Logarithmic plot of the ratio $E_1/\lambda$ for the first
eigenvalue $N=1$ as a function of positive $\mu/\lambda$.
The solid line shows theoretical value,
small squares mark computed numerical values.}
\label{eigen1th}
\end{figure}

One can compare the calculated numerical values of the first eigenvalue
$N=1$ with the theoretical approximate value obtained in \cite{rossi}
\begin{equation}
\frac{E_1}{\lambda} =
\frac{\mu^2}{\sqrt{2\pi} \lambda^2} e^{-\frac{\mu^2}{2\lambda^2}}
\left( 1 - \frac{2\lambda^2}{\mu^2}
 + O\left( \frac{\lambda^4}{\mu^4} \right) \right) .
\label{e55}
\end{equation}
In fact, this expression was found there in a perturbative way:
as the average value of the full Hamiltonian for the approximate
eigenfunction -- the parabolic cylinder function. This theoretical
expression is valid for large positive values of $\mu/\lambda$,
but it is not known how large the value has to be. The comparison
shows that for $\mu/\lambda=5$ the relative error is less than $0.01$
and it decreases at larger $\mu/\lambda$. In Fig.~\ref{eigen1th}
the theoretical value (solid line) and found numerical values
(square points) for positive $\mu/\lambda$ are shown
in a logarithmic scale.

\section{An application: calculation of the propagator}

As an application of eigenvalues of the model \Ref{e13} we consider
the calculation of the one-particle state (one-pomeron state)
propagator. By definition, this propagator is
\begin{equation}
P(y)=<1|e^{-Hy}|1>.
\label{e61}
\end{equation}
The propagator as a function of the rapidity $y$ was found numerically
in \cite{braun1} by means of solving of the evolution equation in $y$
for the one-particle state. In \cite{braun1} six cases were studied:

{\bf (1)} $\mu=1$, $\lambda=0.1$;

{\bf (2)} $\mu=1$, $\lambda=1/3$;

{\bf (3)} $\mu=1$, $\lambda=1$;

{\bf (4)} $\mu=0.1$, $\lambda=1$;

{\bf (5)} $\mu=0$, $\lambda=1$;

{\bf (6)} $\mu=-1$, $\lambda=0.1$.

\noindent
To compare the results we calculated the propagator for the same values
of parameters $\mu$ and $\lambda$.

Inserting the expansion \Ref{e48} between the initial and the final
states in \Ref{e61}, one obtains the spectral representation
\begin{equation}
P(y)=\sum_N \frac{\<1|e^{-Hy}|\psi_N\>\<\bpsi_N|1\>}{\<\bpsi_N|\psi_N\>}
=\sum_N \frac{\<1|\psi_N\>e^{-E_N y}\<\bpsi_N|1\>}{\<\bpsi_N|\psi_N\>} .
\label{e62}
\end{equation}
The eigenfunction as a solution of the differential equation
can be written as
\begin{equation}
\psi_N(z)=\sum_{n=0}^{\infty} c^{[N]}_n x^{n+1}
=\sum_{n=0}^{\infty} c^{[N]}_n
\left( \frac{iz}{\sqrt{2}} \right)^{n+1}
\label{e63}
\end{equation}
with coefficients $c^{[N]}_n$ defined by the recurrence relation
\Ref{e39} using the found real eigenvalue. Thus all these
coefficients are real and normalized by $c^{[N]}_0=1$. From this
$\<1|\psi_N\>=i/\sqrt{2}$ is obvious. Taking
$\<\bpsi_N|1\>=-\<1|\psi_N(-z)\>^*=\<1|\psi_N(z)\>^*=-i/\sqrt{2}$
one obtains
\begin{equation}
P(y)= \frac{1}{2} \sum_N \frac{e^{-E_N y}}{\<\bpsi_N|\psi_N\>} .
\label{e64}
\end{equation}
The propagator has to satisfy the condition $P(0)=1$ which is
the consequence of completeness of the basis $\psi_N$.
The straightforward calculation gives
\begin{equation}
\<\bpsi_N|\psi_N\>=
\sum_{n=0}^{\infty} \frac{(-1)^n (n+1)!}{2^{n+1}} |c^{[N]}_n|^2 .
\label{e65}
\end{equation}
It is not a norm since it is not positively defined, but further
we refer it to as ''norm'' for brevity.
For $\mu<0$, when the finiteness condition for the Bargmann norm
of eigenfunctions
\begin{equation}
\<\psi_N|\psi_N\>=
\sum_{n=0}^{\infty} \frac{(n+1)!}{2^{n+1}} |c^{[N]}_n|^2
\label{e66}
\end{equation}
can be imposed, the series \Ref{e65} converges absolutely.
For $\mu\geq 0$ the sum \Ref{e65} is also finite, although \Ref{e66}
is not \cite{rossi}. Our numerical calculations confirm this.

There are two sources of inaccuracy in this calculation.
First, we know the eigenvalue with a finite precision, hence
the equation $T_{4}=0$ is fulfilled only approximately.
In correspondence with \Ref{e51}, an eigenfunction
determined using an approximate eigenvalue
has a contribution of the growing function $\psi_4$
with a very little coefficient. This contribution grows very rapidly,
so at sufficiently large $|z|$ the approximate eigenfunction differs
significantly from the exact one. In calculation it manifests itself
in solving of the recurrence relations for $c^{[N]}_n$, where
the cumulation of errors occurs which leads to bad convergence
of \Ref{e65}. Second, one can take only a finite number of terms
in the sum \Ref{e65}. Our calculation shows that the values of ''norms''
depend on the number of terms (we got 1000, 2000, 5000, 10000).

Since \Ref{e63} coincides with the exact eigenfunction for small $|z|$
only, one can regularize the series \Ref{e65} by introducing a cut-off
in $|z|$ into the Bargmann scalar product \Ref{e23}. The factor $(n+1)!$
in the numerator of \Ref{e65} is the Bargmann norm of $|n+1\>$.
The introduction of the cut-off $|z|<L$ into the norm \Ref{e24} leads
to the substitution $(n+1)!$ with the integral
\begin{equation}
\int_0^{L^2}\!dt\  e^{-t}t^{n+1} = \Gamma(n+2) - \Gamma(n+2,L^2) ,
\label{e68}
\end{equation}
where $\Gamma(n,z)$ is the upper incomplete gamma function.
The value $e^{-L^2}L^{2(n+1)}$ is the measure of inaccuracy
for this regularization,
hence one has to choose $L$ as large as possible. Also
it is necessary to take into account the cut-off in the numerator of
\Ref{e62} what changes the factor $1/2$ in \Ref{e64} to $B^2/2$, where
\begin{equation}
B=\Gamma(2) - \Gamma(2,L^2) = 1-e^{-L^2}(1+L^2) .
\label{e69}
\end{equation}
Note that in all cases considered below the effect of the last
substitution is negligible (the difference is less than $10^{-10}$).

By this mean we computed ''norms'' $\<\bpsi_N|\psi_N\>$ for
the first $N=1\dots 10$ eigenfunctions in the aforementioned
cases {\bf (2)}--{\bf (5)} and for the first $N=1\dots 6$
eigenfunctions in the case {\bf (6)}, taking 10000 terms in \Ref{e65}.
The used regularization effectively suppresses the contribution
of higher powers $n$, so the sums of 5000 and 10000 terms
practically coincide. We checked the independence of ''norms''
on the cut-off, varying $L^2$ within wide limits. We have begun
with small $L^2=10$ and were sequentially enlarging $L^2$, checking
that the ''norm'' does not change with the relative precision at least
$0.001$. Starting with some value of $L^2$ the regularized ''norms''
became anomalously large. Then we restored the previous value
of $L^2$ and used it for calculating of ''norms'' in propagator
\Ref{e64}. Here are the values of the cut-off $L^2$ for which
''norms'' do not change in the considered cases:

{\bf (2)} - $L^2$ from 25 to 100,
{\bf (3)} - $L^2$ from 25 to 1200,

{\bf (4)} and {\bf (5)} - $L^2$ from 25 to 4000,
{\bf (6)} - $L^2$ from 10 to 30.

\noindent
The condition $P(0)=1$, which is the consequence of the completeness
condition, is fulfilled with an error $0.0025$ in the case {\bf (2)}
and less than $10^{-5}$ in the cases {\bf (3)}--{\bf (6)}.
One can see that the larger the value of $\mu/\lambda$,
the smaller the interval of $|z|$ for which the approximated
eigenfunctions are in good coincidence with the exact ones.

The calculation shows that not all ''norms'' are positive. In the
considered cases their signs alternate for sequential $N$.
Perhaps, it can be explained by the behaviour of eigenfunctions
(see Fig.~\ref{functions4}) on the real axis $x$. For all solutions
$\psi'(0)=1$ is implied. The function $\psi_N(x)$ has $N$ zeroes,
including $x=0$, and its asymptotics at large $x>0$ is constant,
so the sign of this constant is $(-1)^{N-1}$. At $x\to -\infty$ 
the function grows most rapidly, as $\psi_N(x)\sim\exp(x^2+\beta x)/x$,
and has the negative sign. Hence, the main contribution to the ''norm''
\Ref{e46} comes from the real axis $x$ and the sign of this
contribution, if the common ''-'' is taken into account,
coincides with the sign of the constant.

However, the case {\bf (1)} with large positive $\mu/\lambda$
differs radically from the others. Here it is impossible
to select an interval of $L^2$ for which all values of ''norms''
$\<\bpsi_N|\psi_N\>$ do not depend on the cut-off. We considered
the condition $P(0)=1$ and the alternation of signs of ''norms''
as a criterion of success. We tried many variants from $L^2=0.5$
to $L^2=100$, but we can not find any acceptable result.
For the case {\bf (1)} we also tried to use
the resolution of the identity, analogous to \Ref{e48}
but induced by the scalar product \Ref{e49} with the cut-off $\xi<L$
for the integration over $\xi$. We could reliably find only the first
coefficient $1/(\psi_1|\psi_1)$ of the expansion, the others depend on
the cut-off significantly.

Thus our approach to the calculation of propagator, based on the power
series expansion of eigenfunctions, does not allow us to find the
reliable answer in the case of large positive $\mu/\lambda$.
It is worth to note that just in the case of large positive
$\mu/\lambda$ analytical methods of \cite{rossi} work well.

\begin{figure}[h]
\hskip 4.7cm
\epsfig{file=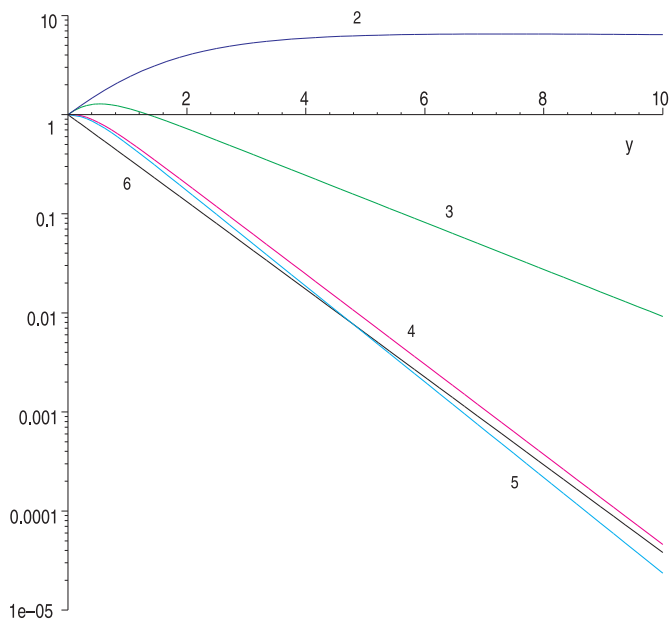,scale=0.6}
\caption{Full propagators for different values $\mu$ and $\lambda$
as functions of $y$.
{\bf (2)} - $\mu=1$, $\lambda=1/3$; {\bf (3)} - $\mu=1$, $\lambda=1$;
{\bf (4)} - $\mu=0.1$, $\lambda=1$; {\bf (5)} - $\mu=0$, $\lambda=1$;
{\bf (6)} - $\mu=-1$, $\lambda=0.1$.}
\label{prop23456}
\end{figure}

In Fig.~\ref{prop23456} plots of the propagator as a function of
rapidity $y$ are shown in a logarithmic scale. The numeration of
curves {\bf 2}--{\bf 5} coincides with that of the considered
cases. In the calculation of \Ref{e64} 10 eigenfunctions for
the cases {\bf (2)}--{\bf (5)} and 6 eigenfunctions for the case
{\bf (6)} were taken into account. The general appearance of
the curves in Fig.~\ref{prop23456} is in full agreement with
the curves in Fig.~1 from \cite{braun1}. The comparison of the
values of the propagators with the original numerical data
for Fig.~1 from \cite{braun1} at integer $y$ from 0 to 20
(plots are shown only for $y<10$) demonstrates that they coincide
with the relative precision less than $0.002$ for all $y>0$.

\begin{figure}[h]
\hskip 5.5cm
\epsfig{file=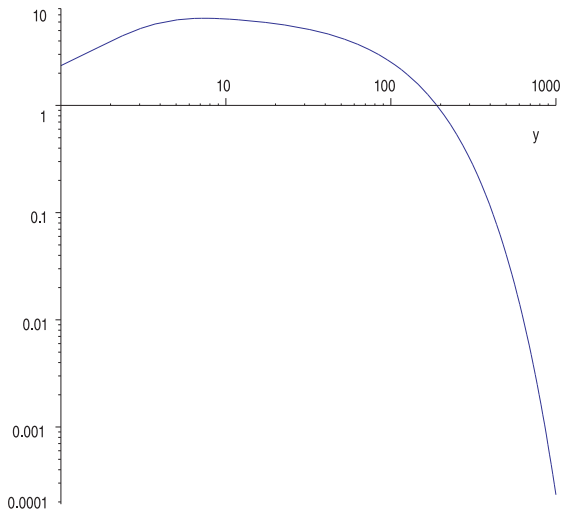,scale=0.5}
\caption{Full propagator for $\mu=1$ and $\lambda=1/3$ as a function
of $y$ with both axes given in the logarithmic scale.}
\label{prop2a}
\end{figure}

In the case {\bf (2)} the propagator grows at small $y$. The physical
propagator tends to zero at $y\to +\infty$, so this growth has to
eventually change to decrease. Theoretically, since the propagator is
defined by \Ref{e64} with positive values $E_N$ (which are growing
with $N$), at sufficiently large $y$ only the decreasing contribution
of $N=1$ remains. In Fig.~\ref{prop2a} the plot of $P(y)$ for $y$
from 1 to 1000 in the case {\bf (2)} is shown with both axes given
in a logarithmic scale. It confirms the decrease of the propagator
at $y>10$.

\section{Conclusions}

In the present work the one-dimensional reggeon model was considered
in its equivalent form of the quantum mechanics in imaginary time.
Since the Hamiltonian of the model is non-Hermitian, the indefinite
scalar product with respect to which the eigenfunctions are orthogonal
had to be introduced. This allows one to write the correct resolution
of the identity and, hence, the spectral representation for
the propagator. The similarity transformation is known to turn
the Hamiltonian into a Hermitian form \cite{jengo}, what establishes
the completeness of the basis of eigenfunctions and reality
of eigenvalues.

The choice of the quantization conditions in this model is not trivial
\cite{rossi}. The condition of finiteness of the norm in the Fock-Bargmann
space can be imposed only for negative values of parameter $\mu$, i.e.
in the area of applicability of the perturbation theory. For $\mu\geq 0$
the asymptotical conditions for eigenfunctions have to be chosen in the
same manner as in the case $\mu<0$, then the energies are real and
positive. In the case $\mu\geq 0$ the asymptotical conditions do not
lead to the finite Bargmann norm, but the non-positive ''norm''
connected with the indefinite scalar product is finite for all cases.

The eigenfunctions equation has the canonical form of the biconfluent
Heun equation for solutions of which the asymptotical conditions are
implied. By resolving these conditions, using the new method
of \cite{numalg}, we can derive the equation which completely defines
the eigenvalues. This equation is transcendental because it contains
the infinite sum of polynomials, so that we can solve it only numerically.
For the chosen sets of parameters of the model (in fact, the only
parameter is the ratio $\mu/\lambda$) we found several eigenvalues
of the energy.

We used the found values of energies for calculation of the pomeron
propagator. In principle, knowing the eigenvalues and eigenfunctions,
one can apply the spectral representation. The problem is that we
express eigenfunctions as power series and their coefficients are
defined with cumulative errors, even if we know eigenvalues with high
precision. These errors lead to bad convergence of the series for the
scalar products appearing as coefficients of the spectral representation.
To provide convergence of the series we introduced a cut-off
into the integration which defines the scalar product. This allows
to calculate the propagator, excluding the case of large positive
$\mu/\lambda$, when the scalar products depend significantly on the
cut-off. So this method gives satisfactory results only for not very
large $\mu/\lambda$, just when the perturbative theory does not work.
In this case we find a full agreement with previous straightforward
numerical calculations \cite{braun1}.

\section{Acknowledgements}

The authors are thankful to N.V. Antonov, M.V. Ioffe, M.V. Komarova,
M.V. Kompaniets, V.N. Kovalenko for very useful discussions.

\section{Appendix. Calculation of the connection factor $T_{4}$}

The following method of calculation of the connection factors
was presented in \cite{numalg}. Let us make a transition from
function $\psi(x)$ following \Ref{e32} with $\alpha=-1$
to $u(x)$, which is a solution of the Heun equation in the normal form
\Ref{e33}. Accordingly, for the asymptotics at $x \to \infty$ one can write
\begin{equation}
u(x) \underset{x \to \infty}{\sim} T_{3} u_3(x) + T_{4} u_4(x) ,
\label{e91}
\end{equation}
where $u_{3,4}(x)=e^{-\frac{\beta x}{2} -\frac{x^2}{2}} \psi_{3,4}(x)$.
Let us denote a Wronskian of two functions by $\mathcal{W}[f,g]=fg' - f'g$.
Forming the Wronskian of both sides of \Ref{e91} with $u_3(x)$ one obtains
\begin{equation}
\mathcal{W}[u,u_3] = T_{4} \mathcal{W}[u_4,u_3].
\label{e92}
\end{equation}
Since $u$, $u_3$ and $u_4$ are solutions of the equation \Ref{e33}, 
expressing their second derivatives from this equation one obtains
$\mathcal{W}'[u,u_3]=0$ and $\mathcal{W}'[u_4,u_3]=0$, hence these
Wronskians are constants and do not depend on $x$. Then to calculate
$\mathcal{W}[u_4,u_3]$ it is enough to use only first terms
of asymptotical expansions \Ref{e311} and after brief calculations
one can obtain that $\mathcal{W}[u_4,u_3]=-\mathcal{W}[u_3,u_4]=-2$.

Introducing auxiliary functions
\begin{equation}
v(x)=\exp\left( \frac{x^2}{4} \right) u(x) ,
\quad v_3 (x)=\exp\left( \frac{x^2}{4} \right) u_3 (x),
\label{e94}
\end{equation}
for their Wronskian one obtains
\begin{equation}
\mathcal{W}[v(x),v_3(x)]=\mathcal{W}[u,u_3] \exp\left( \frac{x^2}{2} \right).
\label{e95}
\end{equation}
It is proposed to find the constant $\mathcal{W}[u,u_3]$ by comparing
asymptotical expansions of the left and right hand sides. Using \Ref{e32}
and \Ref{e94} for the left hand side one obtains
\begin{equation}
\mathcal{W}[v(x),v_3(x)]
=\mathcal{W}[e^{\frac{\beta x}{2} +\frac{x^2}{4}} \hat{u},
e^{-\frac{\beta x}{2} -\frac{x^2}{4}} \psi_3]
= -(x+\beta)\hat{u}\psi_3 + \hat{u}\psi'_3 - \hat{u}'\psi_3 ,
\label{e96}
\end{equation}
where
\begin{equation}
\hat{u}(x) = \exp\left( -\frac{\beta x}{2} \right) u(x) .
\label{e97}
\end{equation}
Function $\hat{u}(x)$ is analytical, its power series is
\begin{equation}
\hat{u}(x) = \sum_{n=0}^{\infty} \hat{c_n} x^{n+1} 
\label{e98}
\end{equation}
and it converges in the entire complex plane $x$.
The coefficients are defined by the recurrence relation
\begin{equation}
n(n+1) \hat{c}_n = \left(-\beta n + \frac{\delta}{2}\right) \hat{c}_{n-1}
-\hat{c}_{n-2} +\beta \hat{c}_{n-3} + \hat{c}_{n-4} ,
\quad \hat{c}_0=1, \quad
( \hat{c}_{n}=0 \mbox{ if } n<0 ).
\label{e99}
\end{equation}
Substituting into \Ref{e96} the power series from \Ref{e98}
and the asymptotical expansion \Ref{e311} for $\psi_3$
one can obtain the asymptotical power series
\begin{equation}
\mathcal{W}[v(x),v_3(x)] \sim \sum_{s=-\infty}^{\infty} g_s x^{s+1} ,
\label{e910}
\end{equation}
where
\begin{equation}
g_s=\sum_{n=0}^{\infty} a^3_n
\left( -\hat{c}_{s+n-1} -\beta \hat{c}_{s+n}
-(s+2n+2)\hat{c}_{s+n+1} \right).
\label{e911}
\end{equation}

In order to obtain the right hand side asymptotical expansion
in \Ref{e95} one should use the Heaviside exponential series
\cite{hardy}
\begin{equation}
e^{t} \sim \sum_{n=-\infty}^{\infty}
\frac{t^{n+\Delta}}{\Gamma (n+1+\Delta)} .
\label{e912}
\end{equation}
This asymptotical expansion is valid for arbitrary
$\Delta\in\mathbb{C}$ and $|\mbox{arg}(t)|<\pi$.
If on the right hand side of \Ref{e95} one expresses
$\mathcal{W}[u,u_3]=\eta_0 + \eta_1$, where $\eta_{0,1}$
are unknown constants, then
\begin{equation}
\mathcal{W}[u,u_3] \exp\left( \frac{x^2}{2} \right) =
\sum_{n=-\infty}^{\infty} \left(
  \eta_0 \frac{ \left( \frac{x^2}{2} \right)^{n+\frac{1}{2}} }
              {\Gamma (n+\frac{3}{2})}
+ \eta_1 \frac{ \left( \frac{x^2}{2} \right)^{n+1} }
              {\Gamma (n+2)} \right),
\label{e913}
\end{equation}
where in the first and second terms $\Delta=1/2$ and $\Delta=1$,
respectively. Thus the sum in the right part of \Ref{e913}
contains all integer powers in $x$. By comparison of the coefficients
of $x^{2\mathcal{N}}$ and $x^{2\mathcal{N}+1}$ ($\mathcal{N}$
is an integer positive number) in \Ref{e910} and \Ref{e913} $\eta_0$
and $\eta_1$ can be defined and then one can obtain
\begin{equation}
\mathcal{W}[u,u_3] =
\frac{\Gamma (\mathcal{N}+\frac{3}{2})}
     {\left( \frac{1}{2} \right)^{\mathcal{N}+\frac{1}{2}}} g_{2\mathcal{N}}
+\frac{\Gamma (\mathcal{N}+2)}
      {\left( \frac{1}{2} \right)^{\mathcal{N}+1}} g_{2\mathcal{N}+1} \ .
\label{e914}
\end{equation}
By construction the right hand side does not depend on $\mathcal{N}$.
Recurrence relations \Ref{e312} and \Ref{e99} define $a^3_n$
and $\hat{c}_n$ as polynomials in $\beta$ and $\delta$,
but the expression for $g_s$ contains an infinite sum.
So that, in the general case, $\mathcal{W}[u,u_3]$ is a transcendent
function of the parameters.

Substituting \Ref{e914} and $\mathcal{W}[u_4,u_3]=-2$ into \Ref{e92},
one obtains \Ref{e53}. The Heaviside expansion in \Ref{e913} can be
used when $|\mbox{arg}(x^2 /2)|<\pi$, therefore the expression \Ref{e53}
for the connection factor is valid only when $-\pi/2<\mbox{arg}(x)<\pi/2$.

\end{document}